\documentclass[showpacs,amsmath,amssymb,aps,showkeys,floatfix,prd,a4paper]{revtex4}

\usepackage[dvips]{graphicx}
\usepackage{dcolumn}
\usepackage{bm}
\usepackage{epsfig}
\usepackage{amsfonts}
\usepackage{amssymb,amscd}

\def\lsim{\raise0.3ex\hbox{$<$\kern-0.75em\raise-1.1ex\hbox{$\sim$}}}

\def\gsim{\raise0.3ex\hbox{$>$\kern-0.75em\raise-1.1ex\hbox{$\sim$}}}

\newcommand{\be}{\begin{equation}}

\newcommand{\ee}{\end{equation}}

\def\beq{\begin{equation}}

\def\eeq{\end{equation}}

\def\beqa{\begin{eqnarray}}

\def\eeqa{\end{eqnarray}}

\newcommand{\rd}{\mbox{\boldmath $\Delta$}}

\newcommand{\ba}{\begin{eqnarray}}

\newcommand{\rr}{\mbox{\boldmath $r$}}

\newcommand{\rb}{\mbox{\boldmath $b$}}

\def\gappeq{\mathrel{\rlap {\raise.5ex\hbox{$>$}}

{\lower.5ex\hbox{$\sim$}}}}

\def\lappeq{\mathrel{\rlap{\raise.5ex\hbox{$<$}}

{\lower.5ex\hbox{$\sim$}}}}

\def\Toprel#1\over#2{\mathrel{\mathop{#2}\limits^{#1}}}

\begin{document}


\title{Exclusive $\rho$ and $J/\Psi$  photoproduction in ultraperipheral $pO$ and $OO$ collisions \\ at energies available at the CERN Large Hadron Collider}

\author{Victor P. {\sc Gon\c{c}alves}}
\email{barros@ufpel.edu.br}
\affiliation{Institut f\"ur Theoretische Physik, Westf\"alische Wilhelms-Universit\"at M\"unster,
Wilhelm-Klemm-Straße 9, D-48149 M\"unster, Germany
}
\affiliation{Institute of Modern Physics, Chinese Academy of Sciences,
  Lanzhou 730000, China}
\affiliation{Institute of Physics and Mathematics, Federal University of Pelotas, \\
  Postal Code 354,  96010-900, Pelotas, RS, Brazil}

\author{Bruno D. {\sc Moreira}}
\email{bduartesm@gmail.com}
\affiliation{Departamento de F\'isica, Universidade do Estado de Santa Catarina, 89219-710 Joinville, SC, Brazil.}

\author{Luana {\sc Santana}}
\email{luana.s@edu.udesc.br}
\affiliation{Departamento de F\'isica, Universidade do Estado de Santa Catarina, 89219-710 Joinville, SC, Brazil.}

\begin{abstract}
Collisions with Oxygen ions at the LHC opens the possibility of investigating the description of the QCD dynamics in an unexplored regime, complementary to those performed in $pp$, $pPb$ and $PbPb$ collisions.
In this paper we estimate the exclusive $\rho$ and $J/\Psi$  photoproduction in ultraperipheral $pO$ and $OO$  collisions at the Large  Hadron Collider using the color dipole formalism and considering distinct models for the dipole - target scattering amplitude and different approaches for the modelling of the vector meson wave function.  We present our predictions for the rapidity and transverse momentum  distributions, as well as for the total cross sections considering the rapidity ranges covered by the ALICE and LHCb detectors. Such results indicate that a future experimental analysis of these final states is feasible and that its study can be useful to improve our understanding of the QCD dynamics at high energies.
\end{abstract}



\maketitle

\vspace{1cm}

\section{Introduction}
The Large Hadron Collider (LHC) recently started its run 3 with the great expectation of new and more precise experimental data that will improve our understanding of the standard model as well as will allow us to search for New Physics in not yet explored kinematical regions. For this new run, the LHC physics programme provides data - taking campaigns with various colliding systems with distinct  center - of - mass energies and characterized by different integrated luminosities \cite{Begel:2022kwp}. In particular, proton - oxygen ($pO$) and oxygen - oxygen ($OO$) collisions  are expected to be studied in the forthcoming years, which will achieve large integrated luminosities in modest running times \cite{Citron:2018lsq}. As summarized in Ref. \cite{Brewer:2021kiv}, these collisions will provide unique  opportunities,  due to the fact that they will allow us to investigate in  detail the transition regime of several phenomena that are expected to have different behaviours  in $PbPb$ and $pp$ collisions \cite{Sievert:2019zjr} as well as they  will also provide valuable data  for cosmic ray modelling \cite{Dembinski:2020dam}. Moreover, the study of $pO$ collisions will also provide important constraints in the nuclear dependence of the parton distribution functions and in the small - $x$ physics (See e.g. Ref. \cite{Paakkinen:2021jjp}).

In this paper we will investigate the exclusive vector meson photoproduction in ultraperipheral $pO$ and $OO$ collisions as a probe of the QCD dynamics at high energies. In these collisions the impact parameter is larger than the sum of the radii of the incident particles, making the photon - induced processes to be dominant \cite{upc}. Our goal is to complement the previous studies performed for $pp$, $pPb$, $PbPb$ and $XeXe$ collisions and present the predictions for the rapidity and transverse momentum distributions, derived considering distinct approaches for the description of the dipole - target interaction and for the treatment of the vector meson wave function (See, e.g. Refs. \cite{run2,Diego1, Diego2}). Our study is strongly motivated by the fact that the saturation scale $Q_s$, which characterizes the transition line between the linear and non - linear regimes of the QCD dynamics \cite{hdqcd}, is predicted to be energy and atomic number dependent. As a consequence, oxygen collisions provide the opportunity to probe the saturation effects in  a new kinematical regime. In what follows, we will focus on the $\rho$ and $J/\Psi$ photoproduction which, as demonstrated in Ref. \cite{run2}, map different configurations of the dipole size and probe distinct regimes of the QCD dynamics. We will present predictions for $pO$ collisions at $\sqrt{s} = 9$ TeV and $OO$ collisions at $\sqrt{s} = 5.52$ TeV, which are possible values for the future runs, but results for other configurations can be provided under request.
{ It is important to emphasize that distinctly from the $J/\Psi$ photoproduction, where the $J/\Psi$ mass provides the hard scale that justifies the perturbative calculation, the analysis of the $\rho$ photoproduction using the dipole approach is a theme of debate in literature. For low energies, $Q^2 \approx 0$ and small values of squared momentum transfer $t$, the $\rho$ production is  dominated by large dipoles, where non - perturbative contributions become non - negligible. However, if non - linear effects are taken into account, another scale is present in the process: the saturation scale $Q_s$, which increases with the energy and with the atomic mass number. The possibility that $Q_s$ provides the hard scale to justify the calculation of the $\rho$ photoproduction has been considered in the literature \cite{Diego1,Diego2,vicmag,Goncalves:2010ci,SampaiodosSantos:2014qtt,Ahmady:2016ujw,Xie:2016ino}, with the predictions in reasonable agreement with the HERA and ALICE data.  Motivated by these previous studies and by the possibility of testing this strong assumption in future runs of the LHC, we will present here the dipole predictions for the $\rho$ photoproduction in $pO$ and $OO$  collisions.}

\begin{figure}[t]
\includegraphics[scale=0.4]{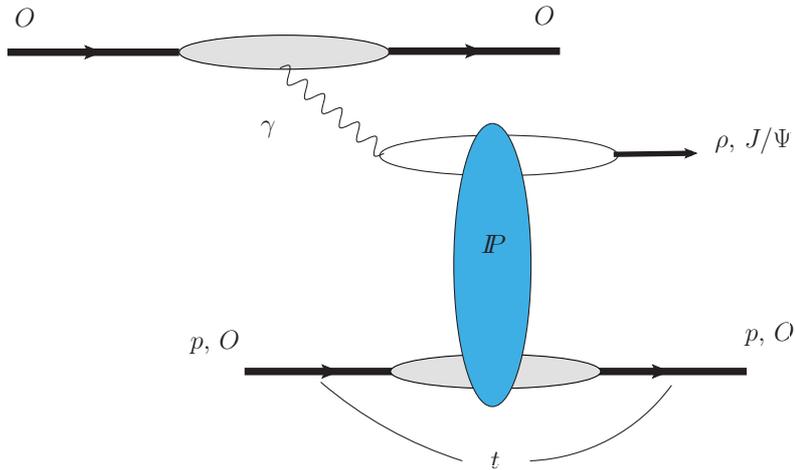}
\caption{Exclusive vector meson photoproduction in ultraperipheral $pO$ and $OO$ collisions.}
\label{Fig:diagram}
\end{figure}

This paper is organized as follows. In the next Section we will present a brief review of the color dipole formalism for the exclusive vector meson production and the models used as input in our calculations will be discussed. Our predictions for the vector meson photoproduction in $pO$ and $OO$ collisions will be presented in Section \ref{Sec:res} and the main conclusions will be summarized in Section \ref{Sec:sum}.

\section{Formalism}

Over the last two decades, the study of the vector meson photoproduction in hadronic colliders, proposed originally in \cite{klein,gluon,Frankfurt:2001db},  became a reality and  it is currently considered the most promising process to improve our understanding of the QCD dynamics and to probe the the transverse spatial distributions of  gluons in the target \cite{Klein:2020nvu}. As the description of this process using the dipole formalism has been largely discussed in the literature (See, e.g. Refs. \cite{upc,Klein:2020nvu}), in what follows we only will present a brief review of the main formulae, needed for the calculation of the observables presented in this paper. The exclusive vector meson photoproduction in $pO$ and $OO$ collisions is represented in Fig. \ref{Fig:diagram} and the associated differential cross section, for a given center - of - mass energy $\sqrt{s}$, can be expressed as follows \footnote{For a recent discussion about the impact of the quantum mechanical interference between these two processes at the amplitude level see, e.g. Ref. \cite{Mantysaari:2022sux}.}
\begin{eqnarray}
\frac{d\sigma \,\left[h_1 + h_2 \rightarrow   h_1 \otimes V \otimes h_2\right]}{dY\,dt}  =  \left[n_{h_1}(\omega)\,\frac{d\sigma}{dt}(\gamma h_2 
\rightarrow V \otimes h_2)\right]_{\omega_L} 
 +  \left[n_{h_2}(\omega)\,\frac{d\sigma}{dt}(\gamma h_1 \rightarrow V \otimes h_1)\right]_{\omega_R}\,
\label{dsigdy}
\end{eqnarray}
where $h_i = p$ or $O$, and the rapidity ($Y$) of the vector meson in the final state is determined by the photon energy $\omega$ in the collider frame and by the mass $M_{V}$ 
of the vector meson [$ Y \propto \ln \, ( \omega/M_{V})$]. Moreover, $d\sigma/dt$ is the differential cross section for the $\gamma h_i \rightarrow V \otimes h_i$ process, with the symbol $\otimes$ representing the presence of a rapidity gap in the final state and $\omega_L \, (\propto e^{-Y})$ and $\omega_R \, (\propto e^{Y})$ denoting  the energies of the photons emitted by the  hadrons $h_1$ and $h_2$, respectively. Furthermore, $n_{h_i}(\omega)$ denotes the  equivalent photon 
spectrum  of the relativistic incident hadron, with the flux of a nucleus   being enhanced by a factor $Z^2$ in comparison to the proton one. 
 As in Refs.  
\cite{run2,Diego1,Diego2} we will assume that the photon flux associated to the proton and to the  nucleus 
can be described by  the Dress - Zeppenfeld  \cite{Dress} and the relativistic point -- like 
charge \cite{upc} models, respectively. In the dipole picture one has that  ${d\sigma}/{dt}$ is given by
\begin{eqnarray}
 \frac{d\sigma}{d{t}}  
= \frac{1}{16\pi} \, |{\cal{A}}_T^{\gamma h \rightarrow V h }(x,\Delta)|^2 \,\,,
\label{sctotal_intt}
\end{eqnarray}
with the scattering amplitude being given by
 \begin{eqnarray}
 {\cal A}_{T}^{\gamma h \rightarrow V h}({x},\Delta)  =  i
\int dz \, d^2\rr \, d^2\rb_h  e^{-i[\rb_h-(1-2z)\rr/2].\rd} 
 \,\, (\Psi^{V*}\Psi)_{T}  \,\,2 {\cal{N}}_{h}({x},\rr,\rb_h) \,\,,
\label{sigmatot2}
\end{eqnarray}
where $\Delta$ denotes the transverse momentum lost by the outgoing hadron (${t} = - \Delta^2$) and $\rb_h$ { is the position of the center of the  $q \bar{q}$  dipole}. Moreover, $(\Psi^{V*}\Psi)_{T}$ denotes the overlap of the transverse photon and vector meson wave functions and  $z$ $(1-z)$ is the longitudinal momentum fractions of the quark (antiquark). For a detailed discussion about the phase factor see, e.g. Refs. \cite{Hatta:2017cte,Mantysaari:2020lhf}. As in previous studies \cite{Diego2,run2},  we will consider the  Gaus - LC (GLC) and Boosted - Gauss (BG) models for the overlap function in order to estimate the impact of the modelling of the vector meson wave function on our predictions. Both models  assume that the vector meson is predominantly a quark-antiquark state and that the spin and polarization structure is the same as in the  photon \cite{dgkp,nnpz,sandapen,KT}, but differ on the description of the scalar part of the wave function (For a more detailed discussion see, e.g.,  Ref. \cite{run2}). 

One has that differential cross section for hadronic collisions, Eq. (\ref{dsigdy}), is proportional to the square of ${\cal{N}}_{h} (x,\rr,\rb_h)$, which is the non-forward scattering  amplitude of a dipole of size $\rr$ on the hadron target for a given value of $x= M^2_V/W^2$, where the $\gamma h$  center - of - mass reaction energy is given by $W = [2 \omega \sqrt{s}]^{1/2}$. Such strong dependence implies that the exclusive vector meson photoproduction in UPCs is sensitive to the description of the QCD dynamics at high energies \cite{vicmag}, which determines ${\cal{N}}_{h}$. Moreover, the analysis of the $t$ - dependence of the predictions can be used to constrain the impact-parameter dependence  of ${\cal{N}}_h$, since $\sqrt{-t}$ and $\rb_h$ are Fourier conjugated variables. It is important to emphasize that recent results indicate  that the Fourier transform of  $d\sigma/dt$ for exclusive processes can be used to obtain the transverse spatial distributions of  gluons in the target and probe the gluon generalized parton distribution (GPD) (See. e.g. Refs. \cite{Hatta:2017cte,Mantysaari:2020lhf}). 
Following Refs. \cite{Diego1,Diego2,run2}, we will assume, for a proton target,  the phenomenological models proposed in Refs. \cite{Kowalski:2003hm,kmw}     that successfully describes the $ep$ HERA data for inclusive and exclusive processes \cite{ipsat4,kmw}. It will allow us to estimate the current theoretical uncertainties associated with the QCD dynamics in the predictions for the exclusive vector meson photoproduction.  The bCGC and IP-Sat models are based on distinct approximations of the   Color Glass Condensate (CGC) formalism \cite{CGC} but predict different behaviors for the linear regime and for transition between the linear and non - linear regimes. One has that the  bCGC model  interpolates two analytical 
solutions of well known evolution equations: the solution of the BFKL equation near  the 
saturation regime and the solution of the  Balitsky-Kovchegov 
equation deeply inside the saturation regime. The underlying assumption is that the interaction between gluonic ladders is taken into account  by bCGC model, with  the saturation boundary being approached via the BFKL equation.  Moreover, this model assumes that the saturation scale  depends on the impact parameter. 
The free parameters were fixed by fitting the HERA data and here we use the updated 
parameters obtained in Ref. \cite{amir}. 
On the other hand, the  IP-Sat model \cite{ipsat4}  assumes an eikonalized form for ${\cal N}_p$  that depends on a gluon distribution evolved via DGLAP equation. This model resums higher twist contributions and, distinctly from the bCGC model,  { the linear regime  is described by the DGLAP evolution \cite{dglap}}. In this work, we employ the parameters  obtained in Ref. \cite{ipsat4}. Finally, for a nuclear target ($A = 16)$, we will  estimate ${\cal{N}}_{O}$  assuming the Glauber-Gribov (GG) formalism~\cite{glauber,gribov,mueller,Armesto:2002ny}, which predicts that \footnote{ { It is important to emphasize that the Glauber - Mueller formalism implies that  the dipole - nucleus scattering amplitude is given by ${\cal{N}}_A(x,\rr,\rb_A) = \left[ 1 - \left(1 - \frac{1}{2}  \, \sigma_{dp}(x,\rr^2) \,T_A(\rb_A)\right)^A \right]$,
 which reduces to an exponentiated form for large $A$ (See e.g. Refs. \cite{Kowalski:2003hm,Kowalski:2008sa}). One has verified that for Oxigen, the predictions derived using these two approximations are almost identical, differing by few percents ($\le 5\%$).   }.}
\begin{eqnarray}
{\cal{N}}_O(x,\rr,\rb_O) =  1 - \exp \left[-\frac{1}{2}  \, \sigma_{dp}(x,\rr^2) \,T_O(\rb_O)\right] \,,
\label{enenuc}
\end{eqnarray}
where the nuclear profile function for the oxygen, $T_O(\rb_O)$, is described by a Woods-Saxon distribution, and $\rb_O$ is the transverse distance from the center of the ion to the center of mass of the $q \bar{q}$  dipole. The dipole-proton cross section, $\sigma_{dp}$, is given by
\begin{eqnarray}
\sigma_{dp}(x,\rr^2) = 2 \int \mathrm{d}^2\rb_p \, {\cal{N}}_p(x,\rr,\rb_p) \,.
\end{eqnarray}
 In our analysis, the  b-CGC and IP-Sat  models for a proton target  will be used as input to estimate ${\cal{N}}_O(x,\rr,\rb_O)$.

{ 
Two comments are in order.  In recent years, several authors have studied the impact of subnucleonic fluctuations on the description of exclusive processes (For a recent review see Ref. \cite{Mantysaari:2020axf})  and obtained that the $t$ - distribution for the incoherent vector meson production is sensitive to  these event-by-event fluctuations of the spatial gluon distribution in the target (See, e.g. Refs. \cite{Mantysaari:2016ykx, Cepila:2017nef,Mantysaari:2017dwh,Krelina:2019gee,Mantysaari:2022sux,Mantysaari:2020lhf}).  
In contrast, for the coherent production, the impact of the geometry evolution on the  rapidity and transverse momentum distributions is small, with the predictions with and without fluctuations being almost identical. As our focus in this paper is in the coherent vector meson photoproduction, we will neglect the geometric evolution of the target in what follows. { Another important comment is that, over the last years, several groups have improved the treatment of exclusive processes in the dipole approach, by estimating higher order corrections for the evolution of the forward dipole - target scattering amplitude and  photon impact factor, as well as by improving the description of the vector meson wave function (See, e.g. Refs. \cite{Boussarie:2016bkq,Lappi:2020ufv,Beuf:2020dxl,Mantysaari:2021ryb,
Mantysaari:2022bsp,Mantysaari:2022kdm}).  Such results indicate that the NLO corrections are numerically important, but their effect can be partially captured when the initial condition for the small-$x$ evolution of the dipole amplitude is fitted to the structure function data. Moreover, the NLO predictions for the vector meson production at HERA are similar to those obtained using the phenomenological dipole models, in particular to those derived using the b-CGC model considered in this paper. Therefore, in principle, we do not expect a large modification of our predictions if the NLO corrections are taken into account. Surely, a more detailed comparison between our results and those derived at NLO is an important next step, which we plan to perform in a future study.}
}

\begin{figure}[t]
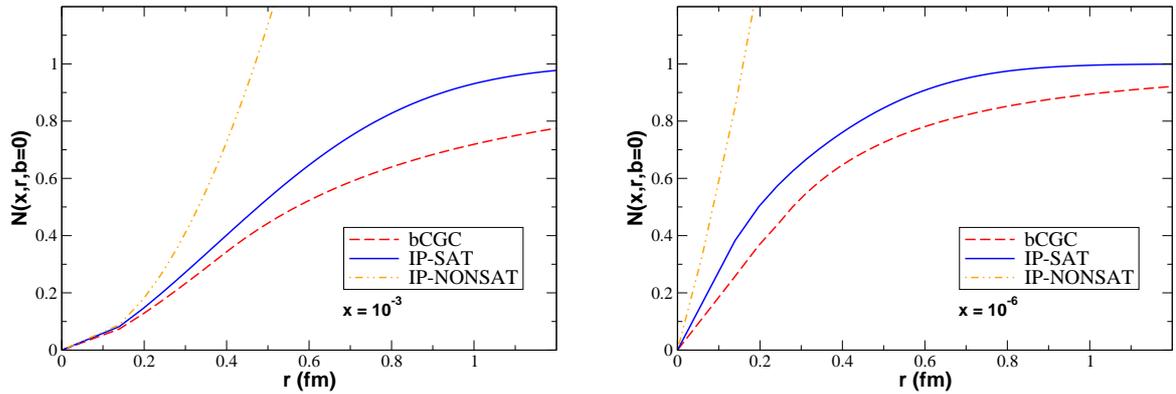

\begin{tabular}{ccc}
\includegraphics[scale=0.31]{n3.eps} & \,\,\,\,\,\,\,\,\,\,\, &
\includegraphics[scale=0.31]{n6.eps}
\end{tabular}
\caption{Dipole-proton scattering amplitude as a function of the  dipole size $r$ for two distinct values of $x$ and  central collisions ($b_p = 0$).}
\label{Fig:enes}
\end{figure}

\begin{figure}[t]
\includegraphics[scale=0.4]{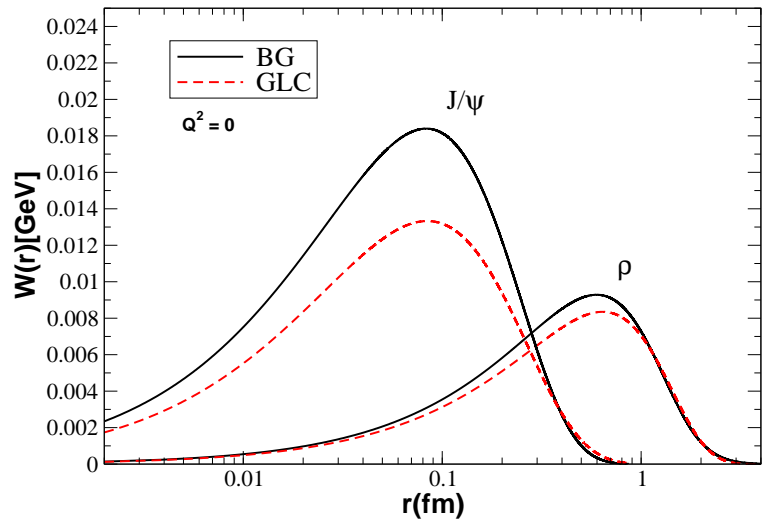}
\caption{Dependence on the dipole size $r$ of the function $W(\rr)$ for the $\rho$ and $J/\Psi$ mesons considering the Boosted Gaussian (BG) and Gaus - LC (GLC)  models for the vector meson wave functions.}
\label{Fig:overlap}
\end{figure}

\begin{figure}[t]
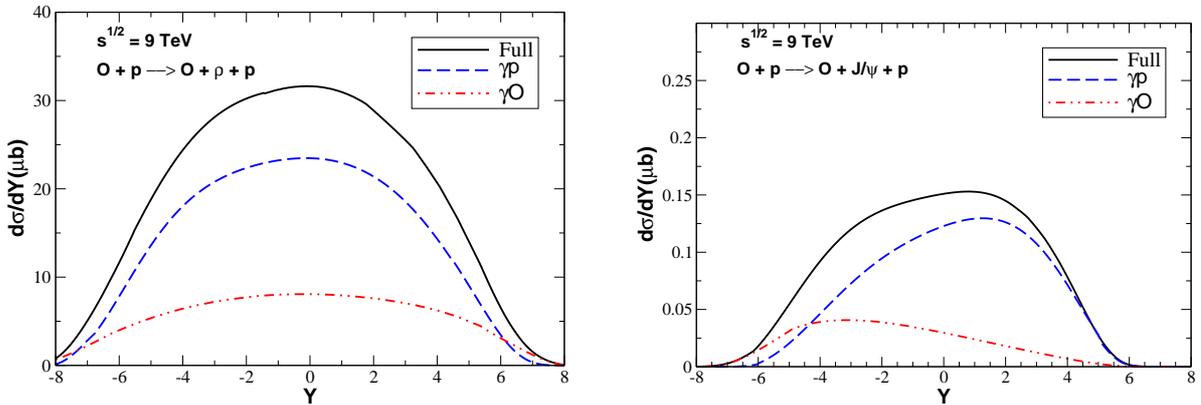

\begin{tabular}{ccc}
\includegraphics[scale=0.31]{pO_compare_rho_UPC.eps} & \,\,\,\,\,\,\,\,\,\,\, &
\includegraphics[scale=0.31]{pO_compare_jpsi_UPC.eps}
\end{tabular}
\caption{Rapidity distributions for the exclusive $\rho$ (left panel) and $J/\Psi$ (right panel) photoproduction in ultraperipheral $pO$ collisions, obtained using  the  Boosted - Gauss model for the vector meson wavefunctions and assuming the bCGC model as input in the calculations of the scattering amplitudes ${\cal{N}}_p$ and ${\cal{N}}_O$. The constributions associated to $\gamma p$ and $\gamma O$ interactions are presented separately as well as the sum of the contributions, denoted Full in the plot.}
\label{Fig:rapidity_partial}
\end{figure}

\section{Results}
\label{Sec:res}

{ Initially, let's discuss the main ingredients of our calculations: the dipole - proton scattering amplitude ${\cal{N}}_p$ and the overlap function $(\Psi^{V*}\Psi)_{T}$. In Fig. \ref{Fig:enes} we present ${\cal{N}}_p$ as a function of the  dipole size $r$ for two distinct values of $x$ and  central collisions ($b_p = 0$). considering the bCGC and IP - SAT models. For comparison, we also present the predictions of the IP-NONSAT model, which is the linear limit of the IP-SAT model, i.e. such a model disregards  the contributions of the non - linear corrections for the QCD dynamics. We have that the description of the linear regime (small - $r$) is distinct in the bCGC and IP-SAT models, as well the transition between the linear and non -- linear  regimes, with the onset of the saturation regime ($  {\cal N}^p \approx 1$) being slower in the case of the bCGC model.  For small dipole sizes and $x = 10^{-3}$, we can observe the different $r$ dependence of the distinct models. In this regime, the   bCGC model predicts that 
${\cal N}^p \propto \rr^{2 \gamma_{eff}}$ for $r^2 \rightarrow 0$, with $\gamma_{eff} \le 1.0$,  while the IP-SAT and IP-NONSAT models predict that ${\cal N}^p \propto \rr^{2} \, xg(x,4/r^2)$.  
On the other hand, for large dipole sizes,  the IP-SAT amplitude has an asymptotic value  larger than the  bCGC one. For $x = 10^{-6}$ we have that the onset of the saturation occurs at smaller values of $r$. The main difference between the bCGC and IP-SAT models is associated with the behavior predicted for the transition between the linear (small-$r$)  and non-linear (large-$r$) regimes of the QCD dynamics. Moreover, it is important to emphasize that the non - linear effects strongly modify the behaviour of    ${\cal{N}}_p$ at large $r$ in comparison to the linear prediction.

In order to analyze the dependence of our predictions on the model assumed for the overlap function $(\Psi^{V*}\Psi)_{T}$,  we present in  Fig. \ref{Fig:overlap} the behavior of the quantity
\begin{eqnarray}
  W(\mbox{\textbf{\textit{r}}}) = 2 \pi r \int_{0}^{1} dz \left[ 
\Psi^{V*}(\mbox{\textbf{\textit{r}}},z) 
  \Psi(\mbox{\textbf{\textit{r}}},z) \right]_T,
 \label{overlap} 
\end{eqnarray}
as a function of  $\rr$ for different vector mesons, considering 
the Boosted Gaussian (BG) and Gaus - LC models for the vector meson function. 
We find that both models predict a peak in the function $W(\mbox{\textbf{\textit{r}}})$. The position of the peak is almost model independent, with the normalization of the GLC model being smaller than the BG one. Moreover,  the peak occurs at larger values of $\rr$ for the $\rho$ meson, while for the $J/\Psi$ it occurs for smaller dipoles. Such a result indicates that for the $J/\Psi$ photoproduction the main contribution for the amplitude comes from dipoles with $r \approx 0.09$ fm, while for the $\rho$ case one has $r \approx 0.7$ fm.  Therefore, studying $\rho$ and $J/\Psi$ photoproduction we are mapping different configurations of the dipole size, which 
probe different regimes of the QCD dynamics, as demonstrated in Fig. \ref{Fig:enes}. It is important to clarify that the predictions for the $\rho$ photoproduction derived using the dipole formalism  can be strongly affected by non-perturbative corrections  not taken into account in our analysis. From Fig. \ref{Fig:overlap} one has that the contribution of dipoles with size larger than 0.7 fm is $\approx 40 \%$ for the $\rho$ case. Therefore, the description of the $\rho$ photoproduction using a perturbative approach is not fully justified. { However, as  the non - linear effects imply that  the contribution of small size dipoles
increases for smaller values of $x$}, one can expect that in the asymptotic regime  $W \rightarrow \infty$ ($x \rightarrow 0$), such a process could be estimated perturbatively. An open question is if the magnitude of $Q_s$ for the energies probed by the current colliders already is large enough to justify the treatment using the dipole approach.  As pointed out in the Introduction,  the dipole predictions provide a reasonable description of the existing HERA and ALICE data. However, we believe that a larger amount of data is needed before  establishing that such an agreement is not accidental. Having it in mind, the comparison of the dipole predictions presented in this paper with the data that can be obtained in future $pO$ collisions will be useful to improve our understanding of the $\rho$ photoproduction.
 }

\begin{figure}[t]
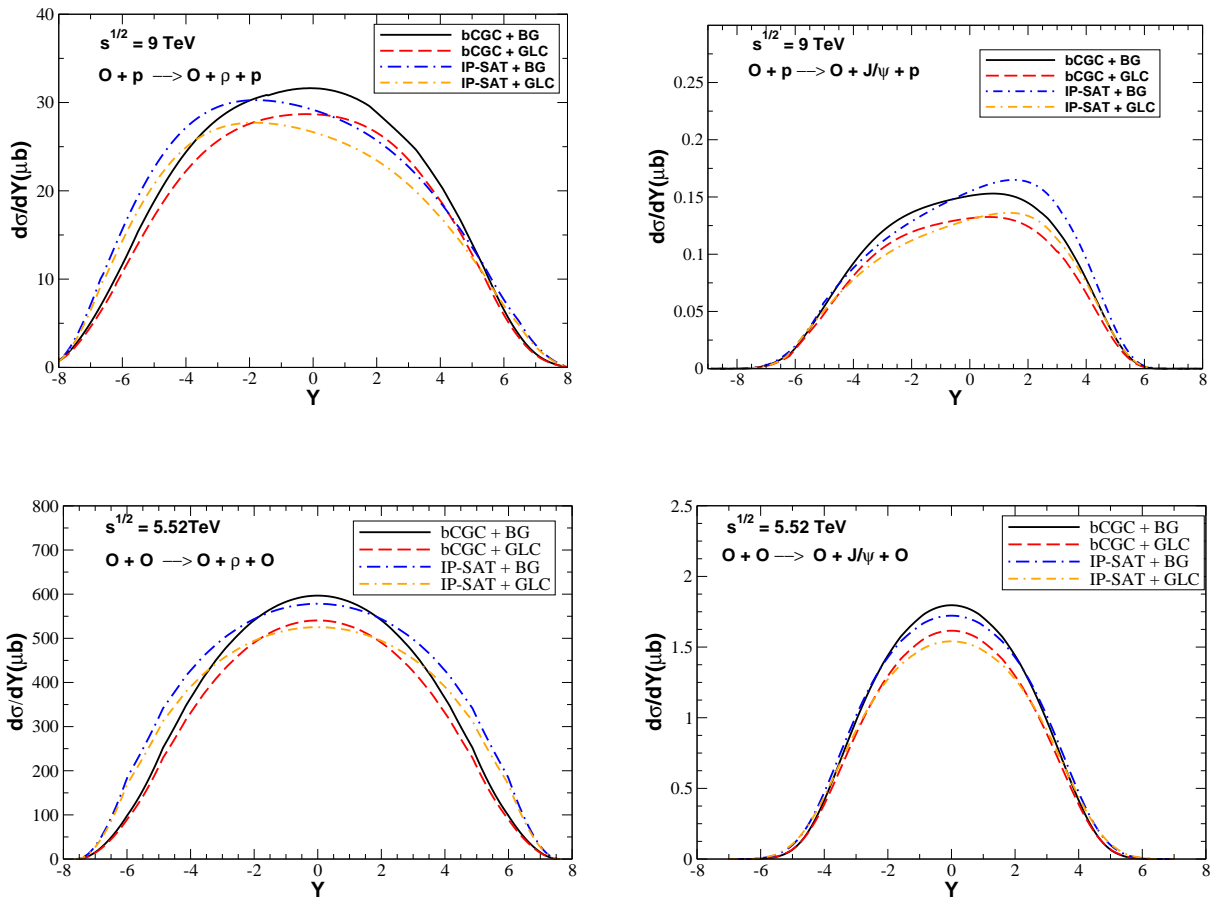

\begin{tabular}{ccc}
\includegraphics[scale=0.31]{NEW_pO_rho_UPC.eps} & \,\,\,\,\,\,\,\,\,\,\, &
\includegraphics[scale=0.31]{NEW_pO_jpsi_UPC.eps} \\
\, & \, & \, \\
\, & \, & \, \\
\, & \, & \, \\
\includegraphics[scale=0.31]{OO_rho_UPC.eps}  & \,\,\,\,\,\,\,\,\,\,\, &
\includegraphics[scale=0.31]{OO_jpsi_UPC.eps}  
\end{tabular}
\caption{Rapidity distributions for the exclusive $\rho$ (left panels) and $J/\Psi$ (right panels) photoproduction in ultraperipheral $pO$ (upper panels) and $OO$ (lower panels) collisions considering  different models for the overlap function and for the description of the dipole - target cross sections.}
\label{Fig:rapidity}
\end{figure}

{ Let's now discuss the exclusive $\rho$ and $J/\Psi$ photoproduction in $pO$ collisions.  For a fixed rapidity, the photon fluxes for the proton and oxygen will be determined by $\omega_L$ and $\omega_R$, respectively, i.e. will probe the photon fluxes in different energy ranges. Moreover, the distribution will be given by the cross sections for $\gamma p$ and $\gamma O$ interactions. In $\gamma p$ interactions the photon comes from the oxygen, with the photon flux being proportional to $Z^2$ ($Z = 8$), and the photoproduction cross section depends on the square of ${\cal{N}}_p$. In $\gamma O$ interactions the photon comes from the proton and the photoproduction cross section is determined by the square of ${\cal{N}}_O$. As a consequence, asymmetric rapidity distributions are expected in $pO$ collisions.  Moreover, one has that the shape of the distribution will be dependent on the model assumed for the dipole - target scattering amplitude and the vector meson that is produced. Such dependencies are expected since the distribution for a given rapidity $Y$ is determined by the product between the photon flux for the photon energy $\omega = M_V/2 \exp(+ Y)$   and the  photon - target cross section for the corresponding photon - target center - of - mass energy $W = \sqrt{2 \omega \sqrt{s} }$. As the photon energy is dependent on the meson mass and the increasing of the cross section with the energy is distinct for the different  models assumed for the overlap function and dipole - target scattering amplitude considered in our analysis, we expect that the shape of the distribution will be different for $\rho$ and $J/\Psi$ production}. Such expectations are confirmed by the results shown in Fig. \ref{Fig:rapidity_partial} and \ref{Fig:rapidity} (upper panels). 

In  Fig. \ref{Fig:rapidity_partial} the predictions for the rapidity distributions  for the exclusive $\rho$ (left panel) and $J/\Psi$ (right panel) photoproduction in ultraperipheral $pO$ collisions, { derived in center - of - mass frame and} obtained by integrating over $t$ the Eq.(\ref{dsigdy}). For this first analysis, we will use  the  Boosted - Gauss model for the vector meson wavefunctions and assume the bCGC model as input in the calculations of the scattering amplitudes ${\cal{N}}_p$ and ${\cal{N}}_O$. {The $O$ beam is assumed to be moving from negative to positive rapidities} and we will present separately the contributions associated to $\gamma p$ and $\gamma O$ interactions as well as the sum of the contributions, denoted Full in the plot. One has that the contribution of $\gamma O$ interactions is non - negligible and can dominate for some values of rapidity, in contrast with the results obtained for $pPb$ collisions, where the $\gamma Pb$ contribution can be disregarded. Such difference arise due to two aspects: (a)  the $Z^2$ enhancement in the photon flux for the oxygen is not a large number, and (b) the maximum energy of the emitted photons by a particle is inversely proportional to its radius, which implies that larger photon - ion center - of - mass energies are reached in ultraperipheral $pO$ collisions in comparison to $pPb$ collisions, which is important when the cross sections increase with the energy as in the process under analysis.

\begin{table}[t]
\begin{center}
\begin{tabular}{||l |r| r|| r| r|| }
\hline
 \bf{ALICE} ($-2.5 \le Y \le 2.5$)&{\bf $\rho$ in $OO$ collisions} & {\bf $\rho$ in $pO$ collisions}  & {\bf $J/\psi$ in $OO$ collisions} & {\bf $J/\psi$ in $pO$ collisions} \\
\hline
 bCGC + BG&  2837.62& 152.70 \,\,\,\,\,\, &  8.06 \,\,\,\,\,& 0.73 \,\,\,\,\,\,\,\\
 \hline
 bCGC + GLC& 2571.39 & 139.24\,\,\,\,\,\,\,\, & 7.25\,\,\,\,\,\,\,  & 0.63 \,\,\,\,\,\,\, \\
 \hline
 IP-SAT + BG& 2800.92 & 142.77\,\,\,\,\,\,\,\, & 7.85 \,\,\,\,\,& 0.75 \,\,\,\,\,\,\, \\
\hline
 IP-SAT + GLC& 2545.44 & 130.45 \,\,\,\,\,\,& 7.01\,\,\,\,\,\,\, & 0.63 \,\,\,\,\,\,\, \\
 \hline
\end{tabular}
\caption{Total cross sections (in $\mu$b) for the exclusive $\rho$ and $J/\Psi$ photoproduction in ultraperipheral $pO$ ($\sqrt{s} = 9$ TeV) and $OO$ ($\sqrt{s} = 5.52$ TeV) collisions at the LHC. Predictions for the rapidity range covered by the ALICE detector derived considering different models for the description of the  dipole - hadron interaction and for the modelling of the overlap function. }
\label{Tab:Alice}
\end{center}
\end{table}

\begin{table}[t]
\begin{center}
\begin{tabular}{||l |r| r|| r| r|| }
\hline
 \bf{LHCb} ($2.0 \le Y \le 4.5$)&{\bf $\rho$ in $OO$ collisions} & {\bf $\rho$ in $pO$ ($Op$) collisions}  & {\bf $J/\psi$ in $OO$ collisions} & {\bf $J/\psi$ in $pO$ ($Op$) collisions} \\
 \hline
 bCGC + BG& 1092.85 & 60.20 (67.15) & 2.06\,\,\,\,\,\,\,& 0.27 (0.28) \\
\hline
 bCGC + GLC& 991.52 & 55.33 (61.31) & 1.88 \,\,\,\,\, & 0.23 (0.24) \\
\hline
 IP-SAT + BG& 1189.18 & 53.87 (71.54) & 2.15\,\,\,\,\,\,\, & 0.31 (0.26) \\
\hline
 IP-SAT + GLC& 1083.71 & 49.10 (65.61) & 1.95\,\,\,\,\,\,\, & 0.25 (0.23) \\
 \hline
\end{tabular}
\caption{Total cross sections (in $\mu$b) for the exclusive $\rho$ and $J/\Psi$ photoproduction in ultraperipheral $pO$ ($\sqrt{s} = 9$ TeV) and $OO$ ($\sqrt{s} = 5.52$ TeV) collisions at the LHC. Predictions for the rapidity range covered by the LHCb detector derived considering different models for the description of the  dipole - hadron interaction and for the modelling of the overlap function. }
\label{Tab:LHCb}
\end{center}
\end{table}

\begin{figure}[t]
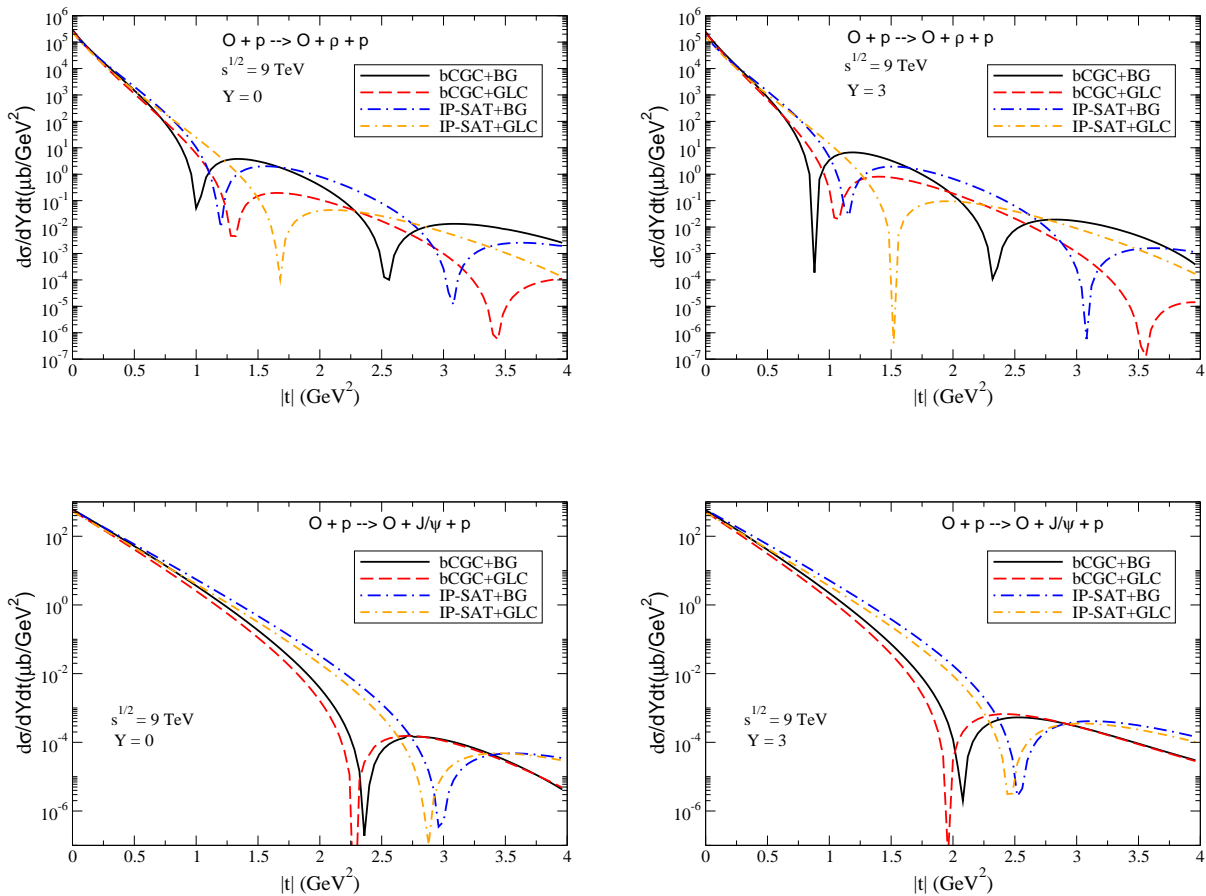

\begin{tabular}{ccc}
\includegraphics[scale=0.31]{dist_t_p_Y0_rho.eps} & \,\,\,\,\,\,\,\,\,\,\, &
\includegraphics[scale=0.31]{dist_t_p_Y3_rho.eps} \\
\, & \, & \, \\
\, & \, & \, \\
\, & \, & \, \\
\includegraphics[scale=0.31]{dist_t_p_Y0_jpsi.eps}  & \,\,\,\,\,\,\,\,\,\,\, &
\includegraphics[scale=0.31]{dist_t_p_Y3_jpsi.eps}  
\end{tabular}
\caption{Predictions for the $t$-distribution of the exclusive $\rho$ (upper panels) and $J/\Psi$ (lower panels) photoproduction in ultraperipheral $pO$ collisions for two values of the meson rapidity: $Y = 0$ (left panels) and $Y = 3$ (right panels).}
\label{Fig:distt_pO}
\end{figure}

The dependence of our predictions for the rapidity distributions on the models assumed for the overlap function and for the description of the dipole - target cross sections are presented in Fig. \ref{Fig:rapidity} considering the exclusive $\rho$ (left panels) and $J/\Psi$ (right panels) photoproduction in ultraperipheral $pO$ (upper panels) and $OO$ (lower panels) collisions. One has that the Gauss - LC (GLC) model for the overlap function diminishes the normalization of the predictions, with a small impact on the shape of the distributions. In contrast, our results for $pO$ collisions (upper panels) indicate that the rapidity distribution is sensitive to the description of the QCD dynamics. 
{ In particular, the asymmetry with relation to $Y = 0$ for the $\rho$ production is different for the bCGC and IP-SAT models due to the distinct behaviours for the transition between the linear (small - $r$) and non - linear (large - $r$) regimes (See Fig. \ref{Fig:enes}). In contrast, for the $J/\Psi$ production, which is dominated by dipoles of small size, one has that both models predict that the maximum of the distribution occurs for positive rapidities. 
Such results indicate that the analyses of both final states can be useful to discriminate between these approaches for the QCD dynamics.} For  $OO$ collisions (lower panels), we have that the main difference between the predictions arises from the model assumed for the overlap function, which is expected due to the fact that the change of bCGC for IP-SAT only affects the argument of the exponential in the Glauber - Mueller model for the dipole - nucleus scattering amplitude [See Eq. (\ref{enenuc})].  
The corresponding predictions for the total cross sections are presented in Tables \ref{Tab:Alice} and \ref{Tab:LHCb} for the rapidity ranges covered by the ALICE and LHCb detectors, respectively. For the LHCb detector, which probes an asymmetric range of rapidities, we present our predictions for $pO$ and $Op$ collisions. We predict values of ${\cal{O}}(mb)$ [${\cal{O}}(\mu b)$] for the $\rho$ [$J/\Psi$] production in $OO$ collisions, with the predictions for $pO$ collisions being smaller by approximately one order of magnitude. Considering the large values of luminosity expected in the case of collisions with Oxygen ions, we can expect a large number of events, which makes the experimental analysis, in principle, feasible.

\begin{figure}[t]
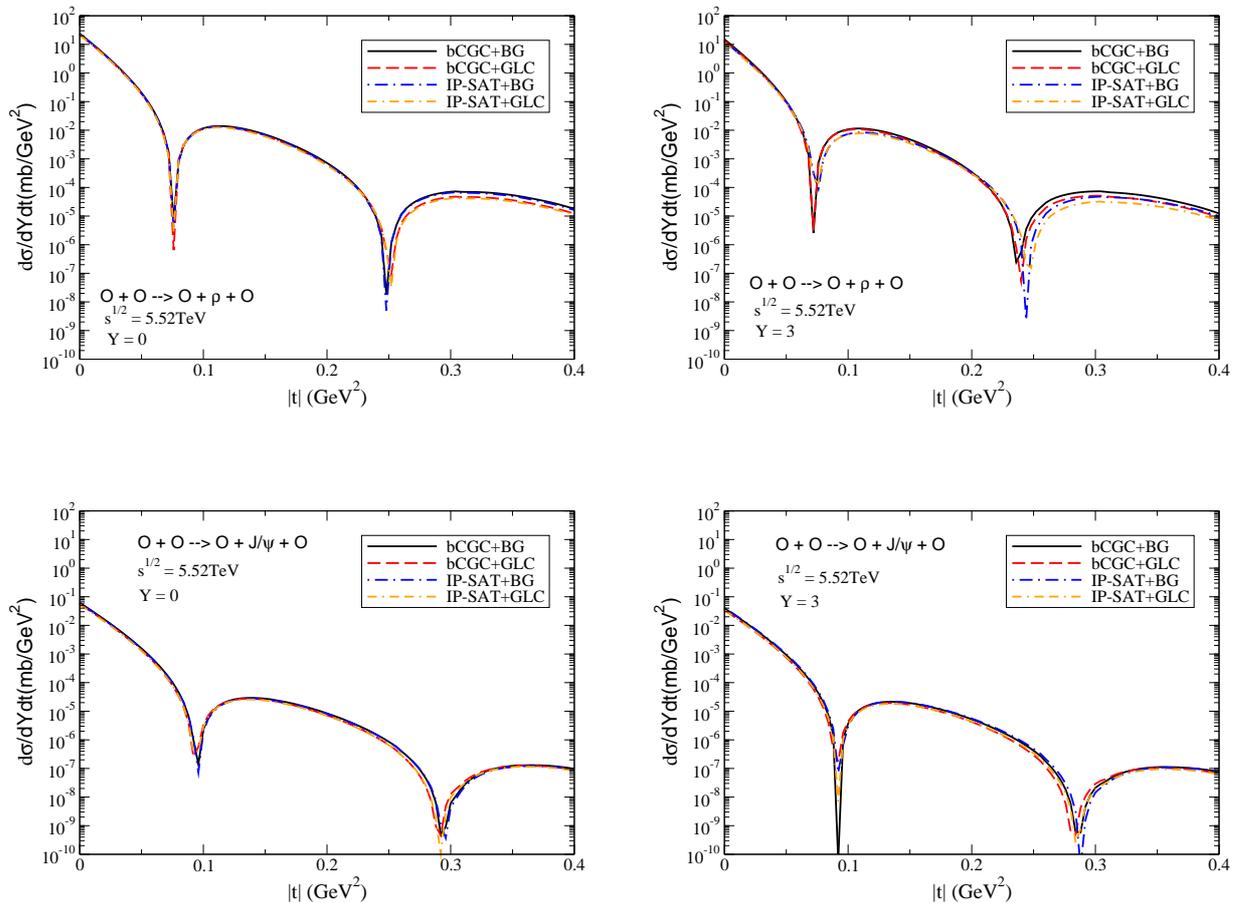

\begin{tabular}{ccc}
\includegraphics[scale=0.31]{dist_OO_Y0_rho2.eps} & \,\,\,\,\,\,\,\,\,\,\, &
\includegraphics[scale=0.31]{dist_OO_Y3_rho2.eps} \\
\, & \, & \, \\
\, & \, & \, \\
\, & \, & \, \\
\includegraphics[scale=0.31]{dist_OO_Y0_jpsi2.eps}  & \,\,\,\,\,\,\,\,\,\,\, &
\includegraphics[scale=0.31]{dist_OO_Y3_jpsi2.eps}  
\end{tabular}
\caption{Predictions for the $t$-distribution of the exclusive $\rho$ (upper panels) and $J/\Psi$ (lower panels) photoproduction in ultraperipheral $OO$ collisions for two values of the meson rapidity: $Y = 0$ (left panels) and $Y = 3$ (right panels).}
\label{Fig:distt_OO}
\end{figure}

The predictions for the $t$ - distribution of the  exclusive vector meson photoproduction in ultraperipheral $pO$ collisions, derived  considering distinct models for ${\cal{N}}_p$ and for the overlap function, are presented in Fig. \ref{Fig:distt_pO} considering two distinct values for the meson rapidity $Y$ { and that  $\gamma p$ interactions are dominant}. The results indicate that the distribution is strongly dependent  on the model considered for the dipole - proton scattering amplitude and for the overlap function. One has that both the bCGC and IP-SAT models  predict dips at large values of $|t|$, with its positions being dependent on the model considered \footnote{{ It is important to emphasize that similar conclusions are also derived if we consider the exclusive vector meson photoproduction at HERA. However, due to experimental restrictions, the H1 and ZEUS Collaborations were not able to measure the coherent $\rho$ and $J/\Psi$ photoproduction at large - $|t|$, where the dips are predicted to occur. As demonstrated e.g. in Refs. \cite{ipsat4,amir}, the current HERA data for low values of $|t|$ are quite well described by the bCGC and IP-SAT models. }}.
 One has that the first dip occurs for smaller values of $|t|$ when the rapidity is increased and for the $\rho$ production. We also have that for the $\rho$ production, the   position of the first dip is strongly dependent on the model assumed for the overlap function. Such is not the case for the $J/\Psi$ production (lower panels), which implies that the analysis of this final state can be useful to constrain the modelling of the QCD dynamics. As already pointed out in the case of the rapidity distribution, these results also indicate that the best way to improve our understanding about the QCD dynamics and vector meson production is the simultaneous analysis of  both final states. The corresponding results for $OO$ collisions are presented in Fig. \ref{Fig:distt_OO}. As in the $pO$ case, the position of the first dip occurs for smaller values of $|t|$ for larger rapidities and in the case of the $\rho$ production. Moreover, one has that the position of the first two dips are similar and become gradually distinct at larger values of $|t|$. In other words, in order to discriminate the treatment of the QCD dynamics at the proton level, we should probe values of $|t| \gtrsim 0.2$ GeV$^2$.

{ A final comment about the feasibility of measurements of the $\rho$ and $J/\Psi$ photoproduction in $pO$ and $OO$ collisions is in order. A current shortcoming to estimate the number of events is associated with the fact that the value of the average luminosity and the number of days available for data acquisition during Run 3 are still the subject of intense debate.  It was discussed in Ref. \cite{Citron:2018lsq} that values of order of $7 \times 10^5 \mu b^{-1}$ could be reached in future runs. Assuming this value and considering our predictions for the total cross sections presented in Tables \ref{Tab:Alice} and \ref{Tab:LHCb}, we predict that the number of $\rho$ ($J/\Psi$) events at ALICE  considering $OO$ collisions will be larger than $17 \times 10^7$ ($49 \times 10^5$). For $pO$ collisions, the predictions are one order of magnitude smaller. As the transverse momentum distributions decrease for large - $t$ (See e.g. Fig. \ref{Fig:distt_pO}), the number of events will be strongly reduced if a cut on the minimum value of $t$ is assumed. As a consequence, a future analysis of the $t$ - distribution in $pO$ collisions will be a challenge for the experimentalists considering the expected luminosity. }

\section{Summary}
\label{Sec:sum}
  The analysis of the exclusive vector meson photoproduction in hadronic colliders is considered one of the most  promising ways  to constrain the QCD dynamics at high energies. In recent years, several studies were performed considering the production of heavy and light vector mesons in $pp$, $pPb$ and $PbPb$ collisions at different center - of - mass energies and/or distinct ranges of rapidity. Study of these different configurations is fundamental to constrain the description of the underlying dynamics since the impact of the non - linear effects are determined by the magnitude of the saturation scale, which is predicted to be energy and atomic number dependent, and is expected to be larger in processes dominated by larger dipole sizes. Such aspects strongly motivate a future experimental analysis of the exclusive $\rho$ and $J/\Psi$ photoproduction in ultraperipheral $pO$ and $OO$ collisions, since these collisions will probe a new kinematical range, complementary to that mapped by previous studies. In particular, they will allow to improve our understanding about the energy and nuclear dependencies of the saturation scale, as well as the description of the linear and non - linear regimes, which are expected to determine the behaviour of the cross sections for the $J/\Psi$ and $\rho$ production, respectively. 
In this paper we have estimated the rapidity and transverse momentum distributions for $pO$ collisions at $\sqrt{s} = 9$ TeV and $OO$ collisions at $\sqrt{s} = 5.52$ TeV, which can be analyzed in forthcoming years. Moreover, we have presented our predictions for the total cross sections considering the rapidity ranges covered by the ALICE and LHCb detectors. Different approaches for the dipole - target scattering amplitude and for the treatment of the overlap function were considered, which allowed to estimate the current theoretical uncertainty on our predictions. Our results indicate that a future experimental analysis is, in principle, feasible.  Considering the results presented in this paper and those in Refs. \cite{Diego1,Diego2}, we strongly motivate the investigation of the $\rho$ and $J/\Psi$ photoproduction in $pO$ and $pPb$ collisions in order to discriminate the distinct approaches for the  QCD dynamics at high energies.

\begin{acknowledgments}
This work was partially supported by CNPq, CAPES, FAPERGS and  INCT-FNA (Process No. 464898/2014-5).  V.P.G. was also partially supported by the CAS President's International Fellowship Initiative (Grant No.  2021VMA0019).
L. Santana  was  partially supported by PROMOP/UDESC.

\end{acknowledgments}

\hspace{1.0cm}


\begin{thebibliography}{99}


\bibitem{Begel:2022kwp}
M.~Begel, S.~Hoeche, M.~Schmitt, H.~W.~Lin, P.~M.~Nadolsky, C.~Royon, Y.~J.~Lee, S.~Mukherjee, J.~Campbell and F.~G.~Celiberto, \textit{et al.}
[arXiv:2209.14872 [hep-ph]].


\bibitem{Citron:2018lsq}
Z.~Citron, A.~Dainese, J.~F.~Grosse-Oetringhaus, J.~M.~Jowett, Y.~J.~Lee, U.~A.~Wiedemann, M.~Winn, A.~Andronic, F.~Bellini and E.~Bruna, \textit{et al.}
CERN Yellow Rep. Monogr. \textbf{7}, 1159-1410 (2019)

\bibitem{Brewer:2021kiv}
J.~Brewer, A.~Mazeliauskas and W.~van der Schee,
[arXiv:2103.01939 [hep-ph]].

\bibitem{Sievert:2019zjr}
M.~D.~Sievert and J.~Noronha-Hostler,
Phys. Rev. C \textbf{100}, no.2, 024904 (2019)


\bibitem{Dembinski:2020dam}
H.~P.~Dembinski, R.~Ulrich and T.~Pierog,
PoS \textbf{ICRC2019}, 235 (2020)

\bibitem{Paakkinen:2021jjp}
P.~Paakkinen,
Phys. Rev. D \textbf{105}, no.3, L031504 (2022)





\bibitem{upc}
C. A. Bertulani and G. Baur, { Phys. Rep.} {\bf 163}, 299 (1988); F.~Krauss, M.~Greiner and G.~Soff,
  Prog.\ Part.\ Nucl.\ Phys.\  {\bf 39}, 503 (1997);
   C.~A. Bertulani, S.~R.~Klein and J.~Nystrand, Ann. Rev. Nucl. Part. Sci. {\bf 55}, 
271 (2005); V.~P.~Goncalves and M.~V.~T.~Machado,
  J.\ Phys.\ G {\bf 32}, 295 (2006);       A.~J.~Baltz {\it et al.},
  Phys.\ Rept.\  {\bf 458}, 1 (2008);       J.~G.~Contreras and J.~D.~Tapia Takaki,
  Int.\ J.\ Mod.\ Phys.\ A {\bf 30}, 1542012 (2015); 
      K.~Akiba {\it et al.} [LHC Forward Physics Working Group],
  J.\ Phys.\ G {\bf 43}, 110201 (2016); S.~R.~Klein and H.~Mantysaari,
Nature Rev. Phys. \textbf{1}, no.11, 662-674 (2019); S.~Klein and P.~Steinberg,
Ann. Rev. Nucl. Part. Sci. \textbf{70}, 323-354 (2020).


\bibitem{run2}
V.~P.~Gon\c{c}alves, M.~V.~T.~Machado, B.~D.~Moreira, F.~S.~Navarra and G.~S.~dos Santos,
Phys. Rev. D \textbf{96}, no.9, 094027 (2017)

\bibitem{Diego1} 
  V.~P.~Goncalves, F.~S.~Navarra and D.~Spiering,
  Phys.\ Lett.\ B {\bf 768}, 299 (2017). 

\bibitem{Diego2} 
  V.~P.~Goncalves, F.~S.~Navarra and D.~Spiering,
Phys. Lett. B \textbf{791}, 299-304 (2019).


  \bibitem{hdqcd} 
  F.~Gelis, E.~Iancu, J.~Jalilian-Marian and R.~Venugopalan,
    Ann.\ Rev.\ Nucl.\ Part.\ Sci.\  {\bf 60}, 463 (2010);
  H.~Weigert,  Prog.\ Part.\ Nucl.\ Phys.\  {\bf 55}, 461 (2005); J.~Jalilian-Marian and Y.~V.~Kovchegov, Prog.\ Part.\ Nucl.\ Phys.\  {\bf 56}, 104 (2006); A.~Morreale and F.~Salazar,
Universe \textbf{7}, no.8, 312 (2021).


 \bibitem{vicmag}
  V.~P.~Goncalves and M.~V.~T.~Machado,
  Eur.\ Phys.\ J.\  C {\bf 40}, 519 (2005).


\bibitem{Goncalves:2010ci}
V.~P.~Goncalves, M.~V.~T.~Machado and A.~R.~Meneses,
Eur. Phys. J. C \textbf{68}, 133-139 (2010)

\bibitem{SampaiodosSantos:2014qtt}
G.~Sampaio dos Santos and M.~V.~T.~Machado,
Phys. Rev. C \textbf{91}, no.2, 025203 (2015)


\bibitem{Ahmady:2016ujw}
M.~Ahmady, R.~Sandapen and N.~Sharma,
Phys. Rev. D \textbf{94}, no.7, 074018 (2016)

\bibitem{Xie:2016ino}
Y.~p.~Xie and X.~Chen,
Eur. Phys. J. C \textbf{76}, no.6, 316 (2016)




\bibitem{klein} S. R. Klein, J. Nystrand,  Phys. Rev. C {\bf 60},
014903 (1999).




\bibitem{gluon}
  V.~P.~Goncalves and C.~A.~Bertulani,
  Phys.\ Rev.\ C {\bf 65}, 054905 (2002).



\bibitem{Frankfurt:2001db}
L.~Frankfurt, M.~Strikman and M.~Zhalov,
Phys. Lett. B \textbf{540}, 220-226 (2002)



\bibitem{Klein:2020nvu}
S.~Klein, D.~Tapia Takaki, J.~Adam, C.~Aidala, A.~Angerami, B.~Audurier, C.~Bertulani, C.~Bierlich, B.~Blok and J.~D.~Brandenburg, \textit{et al.}
[arXiv:2009.03838 [hep-ph]].

\bibitem{Mantysaari:2022sux}
H.~M\"antysaari, F.~Salazar and B.~Schenke,
Phys. Rev. D \textbf{106}, no.7, 074019 (2022)

\bibitem{Dress} M.~Drees and D.~Zeppenfeld, Phys.\ Rev.\ D {\bf
39}, 2536 (1989).

\bibitem{Hatta:2017cte}
Y.~Hatta, B.~W.~Xiao and F.~Yuan,
Phys. Rev. D \textbf{95}, no.11, 114026 (2017)

\bibitem{Mantysaari:2020lhf}
H.~M\"antysaari, K.~Roy, F.~Salazar and B.~Schenke,
Phys. Rev. D \textbf{103}, no.9, 094026 (2021)







\bibitem{dgkp} H.~G.~Dosch, T.~Gousset, G.~Kulzinger and H.~J.~Pirner, Phys. Rev. {\bf D55}, 2602 (1997).
  
\bibitem{nnpz}  
J.~Nemchik, N.~N.~Nikolaev, E.~Predazzi and B.~G.~Zakharov,
  Z.\ Phys.\ C {\bf 75}, 71 (1997)


\bibitem{sandapen}
J.~R.~Forshaw, R.~Sandapen and G.~Shaw,
  Phys.\ Rev.\ D {\bf 69}, 094013 (2004)
  
  
\bibitem{KT}  
 H.~Kowalski and D.~Teaney,
  Phys.\ Rev.\ D {\bf 68}, 114005 (2003)  







\bibitem{Kowalski:2003hm}
H.~Kowalski and D.~Teaney,
Phys. Rev. D \textbf{68}, 114005 (2003).


\bibitem{kmw}
H.~Kowalski, L.~Motyka and G.~Watt,
Phys.\ Rev.\ D {\bf 74}, 074016 (2006).


\bibitem{ipsat4}
A.~H.~Rezaeian, M.~Siddikov, M.~Van de Klundert. and R.~Venugopalan,
Phys.\ Rev.\ D {\bf 87}, 034002 (2013).



\bibitem{CGC}  
J. Jalilian-Marian, A. Kovner, L. McLerran, and H. Weigert, Phys. Rev. D {\bf 55}, 5414 (1997);\\
J. Jalilian-Marian, A. Kovner, and  H.
Weigert, Phys. Rev. D {\bf 59}, 014014 (1999), {\it ibid.} {\bf 59}, 014015 (1999),
{\it ibid.} {\bf 59}  034007 (1999);\\
A. Kovner, J. Guilherme Milhano, and  H. Weigert, Phys. Rev. D {\bf 62}, 114005 (2000);\\
H. Weigert, Nucl. Phys. {\bf A703}, 823 (2002);\\
E. Iancu, A. Leonidov, and L. McLerran,
Nucl.Phys. {\bf A692}, 583 (2001);\\
E. Ferreiro, E. Iancu, A. Leonidov, and L. McLerran,
Nucl. Phys. {\bf A701}, 489 (2002).



\bibitem{amir}
A.~H.~Rezaeian and I.~Schmidt,
Phys.\ Rev.\ D {\bf 88}, 074016 (2013).

\bibitem{dglap}  Yu. Dokshitzer, Sov.\ Phys.\ JETP {\bf 46}, 641 (1977); 
                 V.N. Gribov and L.N. Lipatov, Sov.\ J.\ Nucl.\ Phys.\ {\bf 15}, 438 (1972);
                 G. Altarelli and G. Parisi, Nucl. Phys. B {\bf 126}, 298 (1977).



\bibitem{glauber}
R. J. Glauber, in Lecture in Theoretical Physics, Vol. 1, edited by W. E. Brittin, L. G. Duham (Interscience, New York, 1959).



\bibitem{gribov}
V.~N.~Gribov,
Sov.\ Phys.\ JETP {\bf 29}, 483 (1969); Sov.\ Phys.\ JETP {\bf 30}, 709 (1970).
  

\bibitem{mueller}
A.~H.~Mueller,
Nucl.\ Phys.\ B {\bf 335}, 115 (1990).

\bibitem{Armesto:2002ny}
N.~Armesto,
Eur.\ Phys.\ J.\ C\ {\bf 26}, 35 (2013).



\bibitem{Kowalski:2008sa}
H.~Kowalski, T.~Lappi, C.~Marquet and R.~Venugopalan,
Phys. Rev. C \textbf{78}, 045201 (2008)


\bibitem{Mantysaari:2020axf}
H.~M\"antysaari,
Rept. Prog. Phys. \textbf{83}, no.8, 082201 (2020)


\bibitem{Mantysaari:2016ykx}
H.~M\"antysaari and B.~Schenke,
Phys. Rev. Lett. \textbf{117}, no.5, 052301 (2016)





\bibitem{Mantysaari:2017dwh}
H.~M\"antysaari and B.~Schenke,
Phys. Lett. B \textbf{772}, 832-838 (2017)

\bibitem{Cepila:2017nef}
J.~Cepila, J.~G.~Contreras and M.~Krelina,
Phys. Rev. C \textbf{97}, no.2, 024901 (2018)


\bibitem{Krelina:2019gee}
M.~Krelina, V.~P.~Goncalves and J.~Cepila,
Nucl. Phys. A \textbf{989}, 187-200 (2019)



\bibitem{Boussarie:2016bkq}
R.~Boussarie, A.~V.~Grabovsky, D.~Y.~Ivanov, L.~Szymanowski and S.~Wallon,
Phys. Rev. Lett. \textbf{119} (2017) no.7, 072002





\bibitem{Lappi:2020ufv}
T.~Lappi, H.~M\"antysaari and J.~Penttala,
Phys. Rev. D \textbf{102}, no.5, 054020 (2020)

\bibitem{Beuf:2020dxl}
G.~Beuf, H.~H\"anninen, T.~Lappi and H.~M\"antysaari,
Phys. Rev. D \textbf{102}, 074028 (2020)

\bibitem{Mantysaari:2021ryb}
H.~M\"antysaari and J.~Penttala,
Phys. Lett. B \textbf{823}, 136723 (2021)

\bibitem{Mantysaari:2022bsp}
H.~M\"antysaari and J.~Penttala,
Phys. Rev. D \textbf{105}, no.11, 11 (2022)


\bibitem{Mantysaari:2022kdm}
H.~M\"antysaari and J.~Penttala,
JHEP \textbf{08}, 247 (2022)


\end{thebibliography}
\end{document}